\documentclass{optica-article}
\journal{opticajournal} 
\articletype{Research Article}
\usepackage{lineno}

\usepackage{xcolor}

\begin{document}

%
%

\title{Strong chiral optical force for small chiral molecules based on electric-dipole interactions, inspired by the asymmetrical hydrozoan \textit{Velella velella}}
\author{Robert P. Cameron,\authormark{1,*} Duncan McArthur,\authormark{1} and Alison M. Yao\authormark{1}}
\address{\authormark{1}SUPA and Department of Physics, University of Strathclyde, Glasgow G4 0NG, UK}
\email{\authormark{*}robert.p.cameron@strath.ac.uk} 

\begin{abstract*} 
\noindent Drawing inspiration from a remarkable chiral force found in nature, we show that a static electric field combined with an optical lin$\perp$lin polarization standing wave can exert a chiral optical force on a small chiral molecule that is several orders of magnitude stronger than other chiral optical forces proposed to date, being based on leading electric-dipole interactions rather than relying on weak magnetic-dipole and electric-quadrupole interactions. Our chiral optical force applies to most small chiral molecules, including isotopically chiral molecules, and does not require a specific energy-level structure. Potential applications range from chiral molecular matter-wave interferometry for precision metrology and tests of fundamental physics to the resolution of enantiomers for use in chemistry and biology.
\end{abstract*}

%
%

\section{Introduction}
\label{Introduction}
Interest is growing in the possibility of an optical force that discriminates between the enantiomers of a small\footnote{In this paper, we consider a molecule to be ``small'' if it has a mass of $M\sim 10^2\,\mathrm{Da}$ or less.} chiral molecule \cite{Marichez19a, Kakkanattu21a, Genet22a, Okamoto22a}. Such a force could be used in a wealth of potential applications, including chiral molecular matter-wave interferometry for precision metrology and tests of fundamental physics \cite{Cameron14b, Cameron14c, Stickler21a}, the distillation of chiral molecules in chiral optical lattices as a novel form of matter \cite{Isaule22a}, and the resolution of enantiomers for use in chemistry and biology \cite{Kucirka96a, Canaguier-Durand13a, Cameron14b, Bradshaw15a, Rukhlenko16a, Fang21a, Forbes22a}; a task of particular importance \cite{Lough02a, Gardner05a}. The chiral optical forces proposed to date, however, are extremely weak, as they rely on magnetic-dipole and electric-quadrupole interactions. To the best of our knowledge, no experimental observations of chiral optical forces have been reported for small chiral molecules, although chiral optical forces have been demonstrated for large chiral objects, including chiral liquid crystal microspheres \cite{Cipparrone11a, Hernandez13a, Tkachenko13a, Donato14a, Tkachenko14a, Tkachenko14b, Kravets19a, Kravets19b}, chiral cantilevers \cite{Zhao17a} and chiral gold nanoparticles \cite{Yamanishi22a}.

In this paper, we identify a way forward. Drawing inspiration from a remarkable chiral force found in nature for the asymmetrical hydrozoan \textit{Velella velella} \cite{Alexander03a, Gardner05a}, we show that a static electric field combined with an optical lin$\perp$lin polarization standing wave can exert a chiral optical force on a small chiral molecule that is several orders of magnitude stronger than other chiral optical forces proposed to date, being based on electric-dipole interactions at leading order. Our chiral optical force has the form
\begin{align}
\mathbf{F}&\approx k\mathcal{E}_z\mathcal{E}_y\mathcal{E}_x(\mathtt{A}\alpha_{ZY}\mu_{0X}+\mathtt{B}\alpha_{XZ}\mu_{0Y}+\mathtt{C}\alpha_{YX}\mu_{0Z})\cos(2kZ_0)\hat{\mathbf{z}}, \label{F}
\end{align}
which changes sign if either the electric field or the molecule is inverted, $\mathcal{E}_z\mathcal{E}_y\mathcal{E}_x$ being a chirally sensitive product due to the electric field and $\alpha_{ZY}\mu_{0X}$, $\alpha_{XZ}\mu_{0Y}$, and $\alpha_{YX}\mu_{0Z}$ being chirally sensitive molecular properties. Molecular anisotropy and rotation are accounted for in our theory, with orientational effects playing a crucial role. Our chiral optical force applies to most small chiral molecules, including isotopically chiral molecules (i.e. achiral molecules rendered chiral by isotopic substitution), and does not involve absorption or require a specific energy-level structure. Potential applications identified elsewhere for other chiral optical forces \cite{Kucirka96a, Canaguier-Durand13a, Cameron14b, Cameron14c, Bradshaw15a, Rukhlenko16a, Marichez19a, Fang21a, Kakkanattu21a, Stickler21a, Isaule22a, Genet22a, Okamoto22a, Forbes22a} should be enormously more robust and easier to realise using our chiral optical force instead.

Throughout, we work in vacuum in an inertial frame of reference with time $t$ and position vector $\mathbf{r}=x\hat{\mathbf{x}}+y\hat{\mathbf{y}}+z\hat{\mathbf{z}}$, where $x$, $y$, and $z$ are laboratory-fixed Cartesian coordinates and $\hat{\mathbf{x}}$, $\hat{\mathbf{y}}$, and $\hat{\mathbf{z}}$ are the associated unit vectors. SI units are used, $c$ being the speed of light, $\hbar$ being the reduced Planck constant, $\epsilon_0$ being the electric constant, and $k_B$ being Boltzmann's constant. The Einstein summation convention is to be understood with respect to lower-case Roman indices $a,b,\dots\in\{x,y,z\}$ and uppercase Roman indices $A,B,\dots\in\{X,Y,Z\}$, where $X$, $Y$, and $Z$ are molecule-fixed Cartesian coordinates. Complex quantities are decorated with tildes.

%
%

\section{Origin of our chiral optical force}
\label{Origin of our chiral optical force}
\begin{figure}[h!]
\centering
\includegraphics[width=\textwidth]{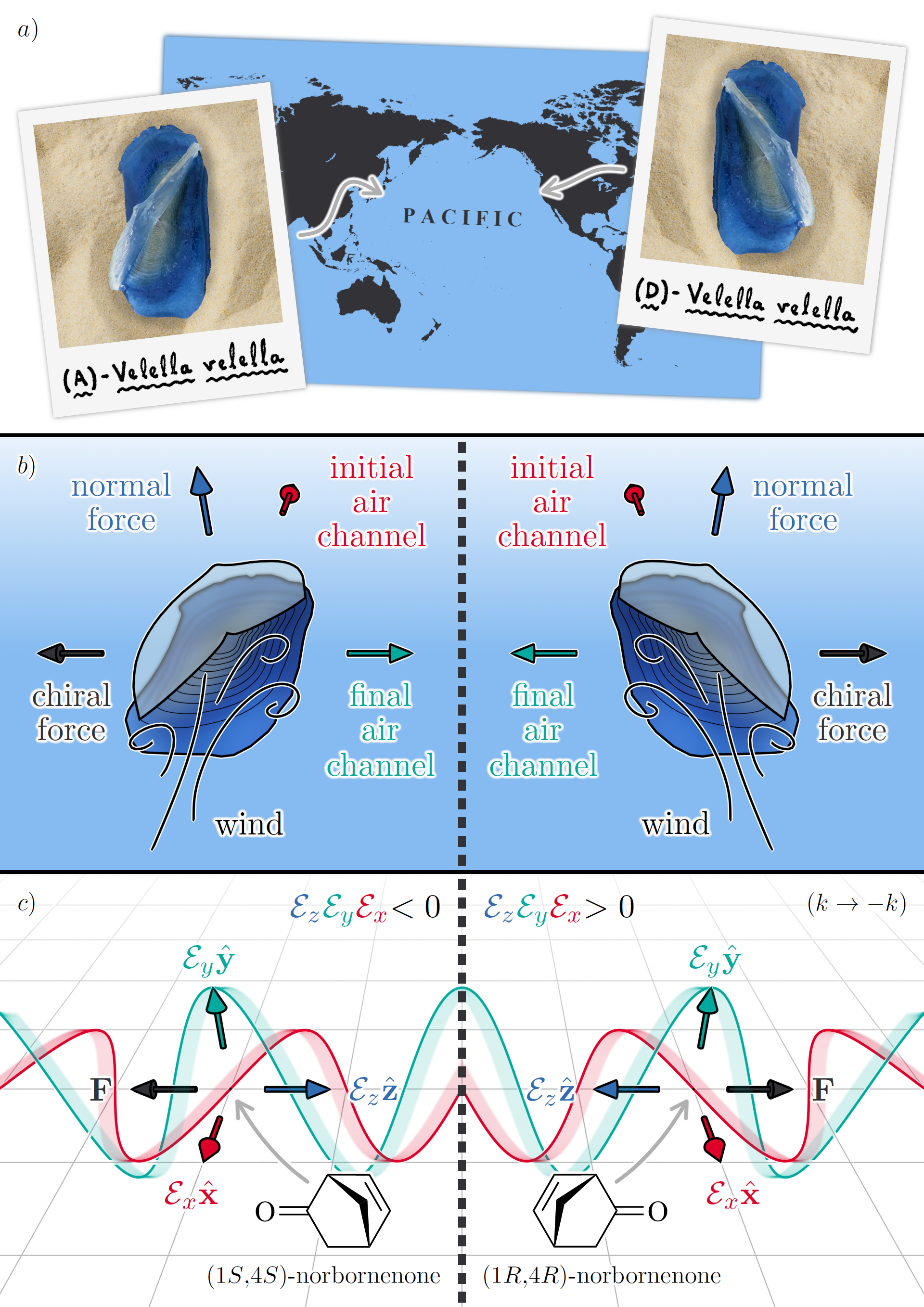}
\caption{\small The two distinct mirror-image forms of \textit{Velella velella} are found on opposite sides of the Pacific a), having been separated by the wind b). Our chiral optical force is constructed by analogy c).}
\label{VelellaVelella}
\end{figure}

This paper was inspired by a remarkable chiral force found in nature. The asymmetrical hydrozoan \textit{Velella velella} lives on the surface of the ocean and has a sail that tilts either antidiagonally or diagonally with respect to a quasi-rectangular base, giving two distinct mirror-image forms. Specimens found on the Japanese side of the Pacific are predominantly of the antidiagonal form whereas those found on the North American side are predominantly of the diagonal form, as illustrated in Fig.~\ref{VelellaVelella} a). It is believed that both forms arise in roughly equal numbers in the center of the Pacific and are then separated by the wind \cite{Alexander03a, Gardner05a}.

The origin of this chiral force can be understood as follows. A \textit{Vellela velella} specimen is orientated by gravity via the normal force. With its sail blown by the wind, the specimen deflects air molecules from the initial air channel into a final air channel with an orthogonal component. The corresponding momentum transfer gives rise to a force with a chiral component. A subtle but crucial ingredient is that the specimen has preferred orientations relative to the wind, due to the quasi-rectangular shape of its base. If all possible orientations were to occur with equal probability instead, the chiral force would vanish on average. The normal force, initial air channel and relevant component of the final air channel form either a left-handed or a right-handed orthogonal triad embodying the chiral sensitivity of the force, as illustrated in Fig.~\ref{VelellaVelella} b).

We construct our chiral optical force by analogy with the above. Instead of a \textit{Velella velella} specimen, we consider a chiral molecule. Instead of gravity, we consider a strong and homogeneous static electric field $\mathbf{E}_0=\mathbf{E}_0(\mathbf{r},t)$ given by
\begin{align}
\mathbf{E}_0&=\mathcal{E}_z\hat{\mathbf{z}}, \label{StaticE}
\end{align}
where $\mathcal{E}_z$ dictates the field's strength and direction. The molecule is (partially) orientated by $\mathbf{E}_0$ via its permanent electric-dipole moment. Instead of the wind, we consider an intense and far off-resonance optical lin$\perp$lin polarization standing wave\footnote{Lin$\perp$lin polarization standing waves are used routinely to laser cool atoms via the Sisyphus effect \cite{Dalibard89a, Cohen-Tannoudji11a}. There is little connection with our chiral optical force.} with complex electric field $\tilde{\mathbf{E}}=\tilde{\mathbf{E}}(\mathbf{r},t)$ given by
\begin{align}
\tilde{\mathbf{E}}&=(\mathcal{E}_y\hat{\mathbf{y}}\mathrm{e}^{\mathrm{i}kz}+\mathrm{i}\mathcal{E}_x\hat{\mathbf{x}}\mathrm{e}^{-\mathrm{i}kz})\mathrm{e}^{-\mathrm{i}\omega t}, \label{ComplexE}
\end{align}
where $\mathcal{E}_y$ dictates the amplitude and phase of the $y$ polarized wave, $\mathcal{E}_x$ dictates the amplitude and phase of the $x$ polarized wave, and $\omega$ is the angular frequency of the waves, $k=\omega/c$ being the associated angular wavenumber. The molecule transfers photons from the $y$ polarized wave to the $x$ polarized wave or \textit{vice versa} \cite{Cohen-Tannoudji11a}. The corresponding momentum transfer gives rise to our chiral optical force. To ensure that the molecule has preferred orientations relative to the $y$ (or $x$) polarized wave, we take its rotational angular momentum to be quantized along the $y$ (or $x$) axis\footnote{In practice, this might be achieved using a static magnetic field, for example.}. If the angular momentum were quantized along the $z$ axis instead, our chiral optical force would vanish. The static electric field $\mathcal{E}_z\hat{\mathbf{z}}$ and the optical polarization vectors $\mathcal{E}_y\hat{\mathbf{y}}$ and $\mathcal{E}_x\hat{\mathbf{x}}$ form either a left-handed ($\mathcal{E}_z\mathcal{E}_y\mathcal{E}_x<0$) or a right-handed ($\mathcal{E}_z\mathcal{E}_y\mathcal{E}_x>0$) orthogonal triad embodying the chiral sensitivity of our force, as illustrated in Fig.~\ref{VelellaVelella} c).

%
%

\section{Derivation of our chiral optical force}
\label{Derivation of our chiral optical force}
In this section, we derive Eq.~(\ref{F}) for our chiral optical force, considering the electromagnetic field defined in Sec. \ref{Origin of our chiral optical force} applied to a small, polar, diamagnetic, chiral molecule in its vibronic ground state with nuclear spins of $0$ or $1/2$ (see Sec. \ref{Rotation degrees of freedom}).

We perform our derivation in two steps. First, we focus on the molecule's vibronic degrees of freedom and derive an expression for the cycle-averaged optical force with the molecule's orientation frozen. This is appropriate assuming the molecule rotates slowly relative to the optical frequency such that its orientation changes little during an optical period, which will usually be the case. Second, we focus on the molecule's rotational degrees of freedom and average the force with respect to the molecule's rotational state, giving our chiral optical force. This two-step approach assumes adiabatic following and ignores rovibronic coupling, which will usually be of little consequence in the vibronic ground state. Throughout, we take the molecule's center of mass position to be held fixed.

\subsection{Vibronic degrees of freedom}
\label{Vibronic degrees of freedom}
We take the instantaneous force $\mathbf{f}=\mathbf{f}(t)$ exerted by the electromagnetic field on the molecule to be given by the Lorentz force law
\begin{align}
\mathbf{f}&=\iiint(\rho\mathbf{E}+\mathbf{J}\times\mathbf{B})\,\mathrm{d}^3\mathbf{r}, \label{F1}
\end{align}
where $\rho=\rho(\mathbf{r},t)$ is the molecule's electric charge density, $\mathbf{J}=\mathbf{J}(\mathbf{r},t)$ is the molecule's electric current density, $\mathbf{E}=\mathbf{E}(\mathbf{r},t)=\mathbf{E}_0+\Re\tilde{\mathbf{E}}$ is the electric field, $\mathbf{B}=\mathbf{B}(\mathbf{r},t)$ is the magnetic field, and the region of integration extends over the entire molecule \cite{Lorentz95a, Griffiths08a}. 

As the molecule is electrically neutral, we have
\begin{align}
\rho&=-\boldsymbol{\nabla}\cdot\mathbf{P} \label{rho} \\
\mathbf{J}&=\frac{\partial\mathbf{P}}{\partial t}+\boldsymbol{\nabla}\times\mathbf{M}, \label{J}
\end{align}
where $\mathbf{P}=\mathbf{P}(\mathbf{r},t)$ is the molecule's polarization field and $\mathbf{M}=\mathbf{M}(\mathbf{r},t)$ is the molecule's magnetization field \cite{Griffiths08a}. Substituting Eqs.~(\ref{rho}) and (\ref{J}) into Eq.~(\ref{F1}) then making use of the triple product $(\boldsymbol{\nabla}\times\mathbf{M})\times\mathbf{B}=(\mathbf{B}\cdot\boldsymbol{\nabla})\mathbf{M}-M_a\boldsymbol{\nabla}B_a$, integration by parts, and the Faraday-Lenz law $\boldsymbol{\nabla}\times\mathbf{E}=-\partial\mathbf{B}/\partial t$, we obtain
\begin{align}
\mathbf{f}&=\iiint(P_a\boldsymbol{\nabla}E_a+M_a\boldsymbol{\nabla}B_a)\,\mathrm{d}^3\mathbf{r}+\frac{\mathrm{d}}{\mathrm{d}t}\iiint\mathbf{P}\times\mathbf{B}\,\mathrm{d}^3\mathbf{r}, \label{F2}
\end{align}
which facilitates a multipolar expansion \cite{Cameron14b}. 

As the molecule is small, we consider only electric-dipole interactions explicitly, taking
\begin{align}
\mathbf{P}&\approx\boldsymbol{\mu}\,\delta^3(\mathbf{r}-\mathbf{R}_0) \label{P} \\
\mathbf{M}&\approx 0, \label{M}
\end{align}
where $\boldsymbol{\mu}=\boldsymbol{\mu}(t)$ is the molecule's electric-dipole moment and $\mathbf{R}_0=X_0\hat{\mathbf{x}}+Y_0\hat{\mathbf{y}}+Z_0\hat{\mathbf{z}}$ is the molecule's center of mass position, with $X_0$, $Y_0$ and $Z_0$ being the Cartesian coordinates \cite{Craig98a}. Substituting Eqs.~(\ref{P}) and (\ref{M}) into Eq.~(\ref{F2}), we obtain
\begin{align}
\mathbf{f}&\approx\mu_a\boldsymbol{\nabla}_{\mathbf{R}_0}E_a(\mathbf{R}_0)+\frac{\mathrm{d}}{\mathrm{d}t}\left[\boldsymbol{\mu}\times\mathbf{B}(\mathbf{R}_0)\right], \label{Fbar1}
\end{align}
which is the well-established result for the Lorentz force experienced by an electric dipole \cite{Gordon73a, Hinds09a}.

As the static electric field $\mathbf{E}_0$ is strong and the standing wave (with electric field $\Re\tilde{\mathbf{E}}$) is intense, we seek an expression for our chiral optical force valid to third order in field components. As Eq.~(\ref{Fbar1}) is already linear in field components, it is sufficient to consider the electric-dipole moment $\boldsymbol{\mu}$ to second order in field components. Remembering that we are considering only electric-dipole interactions explicitly, we take
\begin{align}
\boldsymbol{\mu}&=\boldsymbol{\mu}_0+\Re\tilde{\boldsymbol{\mu}} \label{mu}
\end{align}
with
\begin{align}
\tilde{\mu}_a&\approx\tilde{\alpha}_{ab}(0)E_{0b}+[\tilde{\alpha}_{ab}(\omega)+\tilde{\alpha}_{ab,c}(\omega)E_{0c}]\tilde{E}_b \nonumber \\
&+\frac{1}{2}\tilde{\beta}_{abc}(0;0,0)E_{0b}E_{0c}+\frac{1}{2}\tilde{\beta}_{abc}(2\omega;\omega,\omega)\tilde{E}_b\tilde{E}_c+\frac{1}{2}\tilde{\beta}_{abc}(0;\omega,-\omega)\tilde{E}_b\tilde{E}^\ast_c, \label{complexmu}
\end{align}
where $\boldsymbol{\mu}_0$ is the molecule's permanent electric-dipole moment, $\tilde{\boldsymbol{\mu}}=\tilde{\boldsymbol{\mu}}(t)$ is the molecule's complex induced electric-dipole moment, $\tilde{\alpha}_{ab}=\tilde{\alpha}_{ab}(\Omega)$ is the molecule's complex vibronic polarizability, $\tilde{\alpha}_{ab,c}=\tilde{\alpha}_{ab,c}(\Omega)$\footnote{We have simplified the notation used in \cite{Barron04a} by writing $\tilde{\alpha}_{ab,c}^{(\mu)}\rightarrow\tilde{\alpha}_{ab,c}$.} is a perturbative correction to $\tilde{\alpha}_{ab}$ due to $\mathbf{E}_0$, and $\tilde{\beta}_{abc}=\tilde{\beta}_{abc}(\Omega+\Omega^\prime;\Omega,\Omega^\prime)$ is the molecule's complex vibronic hyperpolarizability, $\Omega$ and $\Omega^\prime$ being angular frequencies \cite{Barron04a}. Eq.~(\ref{complexmu}) consists of a linear static contribution, a linear optical contribution with a linear static correction, a second-order static contribution, a second-order optical harmonic contribution, and a second-order optical rectification contribution, these being all possible first-order and second-order contributions based on electric-dipole interactions. Substituting Eqs.~(\ref{mu}) and (\ref{complexmu}) into Eq.~(\ref{Fbar1}) then taking an average over one optical period, we find that the cycle-averaged optical force $\overline{\mathbf{f}}$ is
\begin{align}
\overline{\mathbf{f}}&=\frac{\omega}{2\pi}\int_0^{2\pi/\omega}\mathbf{f}\,\mathrm{d}t \nonumber \\
&\approx\frac{1}{2}\Re\{[\tilde{\alpha}_{ab}(\omega)+\tilde{\alpha}_{ab,c}(\omega)E_{0c}]\tilde{E}_b(\mathbf{R}_0)\boldsymbol{\nabla}_{\mathbf{R}_0}\tilde{E}^\ast_a(\mathbf{R}_0)\}. \label{Fbar2}
\end{align}
Note that $\tilde{\beta}_{abc}$ does not contribute to Eq.~(\ref{Fbar2}).

The polarizabilities $\tilde{\alpha}_{ab}$ and $\tilde{\alpha}_{ab,c}$ can be partitioned as
\begin{align}
\tilde{\alpha}_{ab}&=\alpha_{ab}(f)+\mathrm{i}\alpha_{ab}(g)-\mathrm{i}[\alpha^\prime_{ab}(f)+\mathrm{i}\alpha^\prime_{ab}(g)] \label{alpha} \\ 
\tilde{\alpha}_{ab,c}&=\alpha_{ab,c}(f)+\mathrm{i}\alpha_{ab,c}(g)-\mathrm{i}[\alpha^\prime_{ab,c}(f)+\mathrm{i}\alpha^\prime_{ab,c}(g)], \label{alphamu}
\end{align}
where $\alpha_{ab}(f)=\alpha_{ba}(f)$ and $\alpha_{ab,c}(f)=\alpha_{ba,c}(f)$ are the time-even, dispersive contributions; $\alpha_{ab}(g)=\alpha_{ba}(g)$ and $\alpha_{ab,c}(g)=\alpha_{ba,c}(g)$ are the time-even, absorptive contributions; $\alpha^\prime_{ab}(f)=-\alpha^\prime_{ba}(f)$ and $\alpha^\prime_{ab,c}(f)=-\alpha^\prime_{ba,c}(f)$ are the time-odd, dispersive contributions; $\alpha^\prime_{ab}(g)=-\alpha^\prime_{ba}(g)$ and $\alpha^\prime_{ab,c}(g)=-\alpha^\prime_{ba,c}(g)$ are the time-odd, dispersive contributions; $f=f_\Omega$ indicates a dispersive lineshape, and $g=g_\Omega$ indicates an absorptive lineshape \cite{Barron04a}. Substituting Eqs.~(\Ref{StaticE}), (\ref{ComplexE}), (\ref{alpha}) and (\ref{alphamu}) into Eq.~(\ref{Fbar2}), we obtain
\begin{align}
\overline{\mathbf{f}}&\approx\overline{\mathbf{f}}(f_\omega)+\overline{\mathbf{f}}(g_\omega)+\overline{\mathbf{f}}^\prime(f_\omega)+\overline{\mathbf{f}}^\prime(g_\omega) \label{Fbar3}
\end{align}
with
\begin{align}
\overline{\mathbf{f}}(f_\omega)&\approx k[\alpha_{yx}(f_\omega)+\alpha_{yx,z}(f_\omega)\mathcal{E}_z]\mathcal{E}_y\mathcal{E}_x\cos(2kZ_0)\hat{\mathbf{z}}, \label{Ff} \\
\overline{\mathbf{f}}(g_\omega)&\approx\frac{1}{2}k\{[\alpha_{yy}(g_\omega)+\alpha_{yy,z}(g_\omega)\mathcal{E}_z]\mathcal{E}_y^2-[\alpha_{xx}(g_\omega)+\alpha_{xx,z}(g_\omega)\mathcal{E}_z]\mathcal{E}_x^2\}\hat{\mathbf{z}}, \label{Fg} \\
\overline{\mathbf{f}}^\prime(f_\omega)&\approx-k[\alpha^\prime_{yx}(f_\omega)+\alpha_{yx,z}^\prime(f_\omega)\mathcal{E}_z]\mathcal{E}_y\mathcal{E}_x\sin(2kZ_0)\hat{\mathbf{z}} , \label{FPrimef} \\
\overline{\mathbf{f}}^\prime(g_\omega)&\approx0, \label{FPrimeg}
\end{align}
where $\overline{\mathbf{f}}(f_\omega)$ is the time-even, dispersive contribution; $\overline{\mathbf{f}}(g_\omega)$ is the time-even, absorptive contribution; $\overline{\mathbf{f}}^\prime(f_\omega)$ is the time-odd, dispersive contribution, and $\overline{\mathbf{f}}^\prime(g_\omega)$ is the time-odd, absorptive contribution. 

As the standing wave is far off-resonance, we neglect absorptive contributions. As the molecule is diamagnetic and thus time-even, we neglect time-odd contributions. Eqs.~(\ref{Fbar3})-(\ref{FPrimeg}) reduce accordingly to
\begin{align}
\overline{\mathbf{f}}&\approx k[\alpha_{yx}(f_\omega)+\alpha_{yx,z}(f_\omega)\mathcal{E}_z]\mathcal{E}_y\mathcal{E}_x\cos(2kZ_0)\hat{\mathbf{z}}. \label{Fbar4}
\end{align}
Eq.~(\ref{Fbar4}) is our final expression for the cycle-averaged optical force with the molecule's orientation frozen.

\subsection{Rotational degrees of freedom}
\label{Rotation degrees of freedom}
As the molecule's nuclear spins are $0$ or $1/2$, we assume that they can be decoupled from the molecule's rotational degrees of freedom using a static magnetic field, for example, and neglect them. Viable isotopes in this regard include ${}^{1}\mathrm{H}$, ${}^{12}\mathrm{C}$, ${}^{13}\mathrm{C}$, ${}^{15}\mathrm{N}$, ${}^{16}\mathrm{O}$, ${}^{18}\mathrm{O}$, and ${}^{19}\mathrm{F}$. If the molecule had nuclear spins of $1$ or higher, they would strongly couple to the molecule's rotational degrees of freedom via electric-quadrupolar interactions \cite{Bragg48a, Bragg49a, Townes75a}, necessitating explicit consideration. 

We take the molecule's rotational state $|\psi\rangle=|\psi(t)\rangle$ to be governed by the Schr\"{o}dinger equation
\begin{align}
\mathrm{i}\hbar\frac{\mathrm{d}|\psi\rangle}{\mathrm{d}t}&=(H^{(0)}+V)|\psi\rangle, \nonumber
\end{align}
where $H^{(0)}$ is the molecule's unperturbed rotational Hamiltonian and $V$ is the molecule's rotational interaction Hamiltonian \cite{Gasiorowicz03a}. $H^{(0)}$ describes the molecule's rotational degrees of freedom in the absence of the electromagnetic field and $V$ accounts for the orientational dependence of the molecule's interaction with the electromagnetic field. We assume that $H^{(0)}\gg|V|$ and treat the interaction perturbatively.

According to basic perturbation theory, the molecule's rotational energy eigenstates $|r\rangle$ and associated energy eigenvalues $E_r$ are given by
\begin{align}
|r\rangle&=|r\rangle^{(0)}+\sum_{q\ne r}\frac{{}^{(0)}\langle q|V|r\rangle^{(0)}}{E_r^{(0)}-E_q^{(0)}}|q\rangle^{(0)}+\dots \label{PerturbedState} \\
E_r&=E_r^{(0)}+{}^{(0)}\langle r|V|r\rangle^{(0)}+\dots, \label{PerturbedEnergy}
\end{align}
where the $|r\rangle^{(0)}$ are the molecule's unperturbed rotational energy eigenstates, the $E^{(0)}_r$ are the associated energy eigenvalues, and we assume that the rotational interaction Hamiltonian $V$ is diagonal in degenerate blocks, with no accidental near degeneracies of importance \cite{Gasiorowicz03a}. The molecule's rotational state $|\psi\rangle$ can be expanded in terms of the $|r\rangle$ and the $E_r$ as
\begin{align}
|\psi \rangle&=\sum_r\tilde{a}_r|\psi_r\rangle \label{RotationalStateI}
\end{align}
with
\begin{align}
|\psi_r\rangle&=\exp(-\mathrm{i}E_r t/\hbar)|r\rangle, \label{RotationalStateII}
\end{align}
where the $|\psi_r\rangle=|\psi_r(t)\rangle$ are the molecule's stationary rotational states and the $\tilde{a}_r$ are the associated probability amplitudes \cite{Gasiorowicz03a}.

Taking the expectation value of Eq.~(\ref{Fbar4}) with respect to Eqs.~(\ref{RotationalStateI}) and (\ref{RotationalStateII}) using Eqs.~(\ref{PerturbedState}) and (\ref{PerturbedEnergy}), we find that the cycle-averaged, rotationally averaged optical force $\langle\overline{\mathbf{f}}\rangle=\langle\overline{\mathbf{f}}\rangle_\psi$ is
\begin{align}
\langle\overline{\mathbf{f}}\rangle&=\langle\psi|\overline{\mathbf{f}}|\psi\rangle \nonumber \\
&\approx k\langle\alpha_{yx}(f_\omega)+\alpha_{yx,z}(f_\omega)\mathcal{E}_z\rangle\mathcal{E}_y\mathcal{E}_x\cos(2kZ_0)\hat{\mathbf{z}} \label{ExpectationForce} 
\end{align}
with
\begin{align}
\langle \alpha_{yx}(f_\omega)+\alpha_{yx,z}(f_\omega)\mathcal{E}_z\rangle&\approx\sum_r\sum_s\tilde{a}^\ast_r\tilde{a}_s\mathrm{e}^{\mathrm{i}(E_r^{(0)}-E_s^{(0)}+{}^{(0)}\langle r|V|r\rangle^{(0)}-{}^{(0)}\langle s|V|s\rangle^{(0)}+\dots)t/\hbar}\nonumber \\
&\times\Bigg[{}^{(0)}\langle r|\alpha_{yx}(f_\omega)+\alpha_{yx,z}(f_\omega)\mathcal{E}_z|s\rangle^{(0)}  \nonumber \\
&+\sum_{q\ne r}\frac{{}^{(0)}\langle r|V|q\rangle^{(0)}{}^{(0)}\langle q|\alpha_{yx}(f_\omega)|s\rangle^{(0)}}{E_r^{(0)}-E_q^{(0)}} \nonumber \\
&+\sum_{q\ne s}\frac{{}^{(0)}\langle r|\alpha_{yx}(f_\omega)|q\rangle^{(0)}{}^{(0)}\langle q|V|s\rangle^{(0)}}{E_s^{(0)}-E_q^{(0)}}+\dots\Bigg]. \label{ExpectationExpression}
\end{align}
The first contribution inside the square brackets in Eq.~(\ref{ExpectationExpression}) describes the force that would be obtained if the molecule rotated freely and the remaining contributions describe perturbative corrections due to the orientational dependence of the molecule's interaction with the electromagnetic field. To evaluate Eqs.~(\ref{ExpectationForce}) and (\ref{ExpectationExpression}), we need an explicit model of the molecule's rotational degrees of freedom.

As the molecule is small, chiral and in its vibronic ground state, we treat it simply as an asymmetric rigid rotor quantized along the $y$ axis\footnote{Quantization along the $x$ axis instead sees our chiral optical force change sign. Quantization along the $z$ axis sees our chiral optical force vanish.}, the laboratory-fixed components $\mu_{0a}$, $\alpha_{ab}(f_\omega)$, and $\alpha_{ab,c}(f_\omega)$ being related to the corresponding molecule-fixed components $\mu_{0A}$, $\alpha_{AB}(f_\omega)$, and $\alpha_{AB,C}(f_\omega)$ via
\begin{align}
\mu_{0a}&=\ell_{aA}\mu_{0A}, \label{MuRotation} \\
\alpha_{ab}(f_\omega)&=\ell_{aA}\ell_{bB}\alpha_{AB}(f_\omega) ,\label{AlphaRotation} \\
\alpha_{ab,c}(f_\omega)&=\ell_{aA}\ell_{bB}\ell_{cC}\alpha_{AB,C}(f_\omega), \label{AlphaMuRotation}
\end{align}
where the $\ell_{aA}$ are direction cosines, as described in Appendix \ref{Asymmetric rigid rotor}.

Recall that we we seek an expression for our chiral optical force valid to third order in field components. As Eq.~(\ref{ExpectationForce}) is already second order in field components at leading order, it is sufficient to consider the rotational interaction Hamiltonian $V$ to first order in field components. Remembering once more that we are considering only electric-dipole interactions explicitly, we take
\begin{align}
V&\approx-\boldsymbol{\mu}_0\cdot\mathbf{E}_0, \label{V}
\end{align}
which is the potential energy of the molecule's permanent electric-dipole moment $\boldsymbol{\mu}_0$ in the static electric field $\mathbf{E}_0$. This form of $V$ is trivially degenerate in diagonal blocks, as the linear Stark effect vanishes for an asymmetric rigid rotor. We have tacitly assumed here that $|-\boldsymbol{\mu}_0\cdot\mathbf{E}_0|\gg|-\alpha_{ab}(f_0)E_{0a}E_{0b}/2-\alpha_{ab}(f_\omega)\tilde{E}_a\tilde{E}_b^\ast/4|$, where $-\alpha_{ab}(f_0)E_{0a}E_{0b}/2-\alpha_{ab}(f_\omega)\tilde{E}_a\tilde{E}_b^\ast/4$ accounts for the orientational dependence of the molecule's polarisable Stark interactions with the electric field to leading order.

\begin{figure}[h!]
\centering
\includegraphics[width=\textwidth]{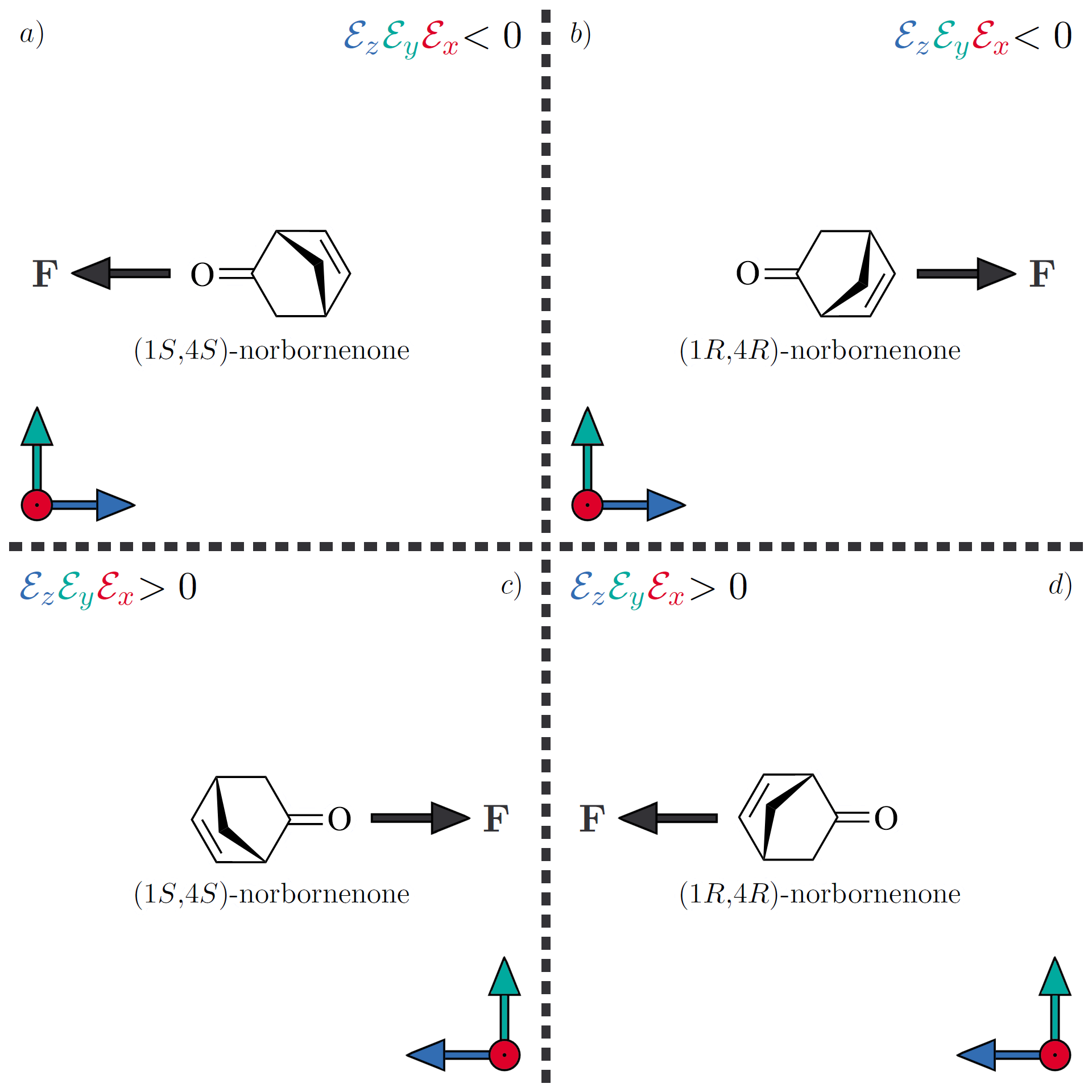}
\caption{\small For a given configuration of the electric field, our chiral optical force has opposite signs for opposite enantiomers in the same rotational state; compare a) with b) and c) with d). For a given enantiomer in a given rotational state, our chiral optical force has opposite signs depending on whether the electric field is in left-handed or a right-handed configuration; compare a) with c) and b) with d).}
\label{Enantioselectivity}
\end{figure}

Let us suppose now that the molecule's rotational state has the stationary\footnote{When the Stark interaction terms $-\alpha_{ab}(f_0)E_{0a}E_{0b}/2-\alpha_{ab}(f_\omega)\tilde{E}_a\tilde{E}_b^\ast/4$ are included in the rotational interaction Hamiltonian $V$, we find that rotational states quantized along the $y$ (or $x$) axis are not necessarily stationary. This limits the validity of our treatment to suitably short interaction times.} form
\begin{align}
|\psi\rangle&=|\psi_{J_\tau,m}\rangle, \label{PsiRotor}
\end{align}
where $J$ and $m$ are rotational quantum numbers and $\tau$ is an energy label, as described in Appendix \ref{Asymmetric rigid rotor}. Substituting Eqs.~(\ref{MuRotation})-(\ref{PsiRotor}) into Eqs.~(\ref{ExpectationForce}) and (\ref{ExpectationExpression}) with Eq.~(\ref{ExpectationExpression}) considered to first order in $V$ then making use of basic symmetry arguments to disregard vanishing contributions, we obtain
\begin{align}
\langle\overline{\mathbf{f}}\rangle&\approx k\mathcal{E}_z\mathcal{E}_y\mathcal{E}_x[\mathtt{a}\alpha_{ZY,X}(f_\omega)+\mathtt{b}\alpha_{XZ,Y}(f_\omega)+\mathtt{c}\alpha_{YX,Z}(f_\omega) \nonumber \\
&+\mathtt{A}\alpha_{ZY}(f_\omega)\mu_{0X}+\mathtt{B}\alpha_{XZ}(f_\omega)\mu_{0Y}+\mathtt{C}\alpha_{YX}(f_\omega)\mu_{0Z}]\cos(2kZ_0)\hat{\mathbf{z}}, \label{ChiralOpticalForceFull}
\end{align}
where $\mathtt{a}$, $\mathtt{b}$, $\mathtt{c}$, $\mathtt{A}$, $\mathtt{B}$, and $\mathtt{C}$ are coefficients that depend on the molecule's rotational state, as described in Appendix \ref{Coefficients}. The $\mathtt{a}$, $\mathtt{b}$, and $\mathtt{c}$ contributions are due to perturbation of the molecule's vibronic degrees of freedom by the static electric field $\mathbf{E}_0$ whereas the $\mathtt{A}$, $\mathtt{B}$, and $\mathtt{C}$ contributions are due to perturbation of the molecule's rotational degrees of freedom. Note that $\langle \alpha_{yx}(f_\omega)\rangle$ does not contribute to Eq.~(\ref{ChiralOpticalForceFull}) at our current level of description. The molecular properties $\alpha_{ZY}(f_\omega)\mu_{0X}$, $\alpha_{XZ}(f_\omega)\mu_{0Y}$, $\alpha_{YX}(f_\omega)\mu_{0Z}$, $\alpha_{ZY,X}(f_\omega)$, $\alpha_{XZ,Y}(f_\omega)$ and $\alpha_{YX,Z}(f_\omega)$ are chirally sensitive, having opposite signs for opposite enantiomers.

As the vibronic excitations of a molecule are usually much more energetic than the rotational excitations, we assume that the $\mathtt{a}$, $\mathtt{b}$, and $\mathtt{c}$ contributions are much smaller than the $\mathtt{A}$, $\mathtt{B}$, and $\mathtt{C}$ contributions and neglect them. Eq.~(\ref{ChiralOpticalForceFull}) reduces accordingly to 
\begin{align}
\mathbf{F}&\approx k\mathcal{E}_z\mathcal{E}_y\mathcal{E}_x(\mathtt{A}\alpha_{ZY}\mu_{0X}+\mathtt{B}\alpha_{XZ}\mu_{0Y}+\mathtt{C}\alpha_{YX}\mu_{0Z})\cos(2kZ_0)\hat{\mathbf{z}}, \nonumber
\end{align}
where we have simplified our notation by writing $\langle\overline{\mathbf{f}}\rangle\rightarrow \mathbf{F}=\mathbf{F}_{J_\tau,m}$ and $\alpha_{AB}(f_\omega)\rightarrow\alpha_{AB}$. This is our chiral optical force, as seen in Eq.~(\ref{F}). 

Our chiral optical force depends on the chirality of the electric field $\mathbf{E}$ through the product $\mathcal{E}_z\mathcal{E}_y\mathcal{E}_x$ and on the chirality of the molecule through the molecular properties $\alpha_{ZY}\mu_{0X}$, $\alpha_{XZ}\mu_{0Y}$ and $\alpha_{YX}\mu_{0Z}$, as illustrated in Fig.~\ref{Enantioselectivity}. To distinguish it from other forces proposed to date, our chiral optical force might be referred to as a ``chiral optical force from $\mathcal{E}_z\mathcal{E}_y\mathcal{E}_x$'' or ``COFFEEE''.

%
%

\section{Numerical results}
\label{Numerical results}
In this section, we present numerical results for our chiral optical force. We focus on a right-handed electric field configuration ($\mathcal{E}_z\mathcal{E}_y\mathcal{E}_x>0$) and single enantiomers, remembering that our chiral optical force changes sign if either the electric field or the molecule is inverted.

For the static electric field, we consider a field strength of $\mathcal{E}_z=10.0\,\mathrm{V}\,\mathrm{mm}^{-1}$, readily achievable using a high voltage source. For the standing wave, we consider field strengths of $\mathcal{E}_y=\mathcal{E}_x=3.16\,\mathrm{kV}\,\mathrm{mm}^{-1}$ (intensity $I=\epsilon_0c\mathcal{E}_y\mathcal{E}_x=2.65\times 10^6\,\mathrm{W}\,\mathrm{cm}^{-2}$) and a near-infrared wavelength of $2\pi/k=1.064\,\mu\mathrm{m}$, readily achievable using a Nd:YAG laser. Much stronger fields are possible, giving rise to even stronger chiral optical forces than those described below. To calculate these forces reliably, however, a non-perturbative treatment of our chiral optical force is required, which is beyond the scope of this paper.

For the molecule, we calculate the rotational constants $A$, $B$, and $C$ and the chirally sensitive properties $\alpha_{ZY}\mu_{0X}$, $\alpha_{XZ}\mu_{0Y}$, and $\alpha_{YX}\mu_{0Z}$ using Gaussian 09, optimizing the molecule's geometry at the DFT B3LYP/6-311++G(d,p) level of theory then employing the DFT B3LYP method with the AUG-cc-pVDZ basis set. Using these values of $A$, $B$, and $C$, we calculate the molecule's unperturbed rotational spectrum as described in Appendix \ref{Asymmetric rigid rotor}. Using this spectrum, we calculate the coefficients $\mathtt{A}$, $\mathtt{B}$, and $\mathtt{C}$ as described in Appendix \ref{Coefficients}.

Eq.~(\ref{F}) for our chiral optical force can be recast as
\begin{align}
\mathbf{F}&\approx-\pmb{\nabla}_{\mathbf{R}_0}U \nonumber
\end{align}
with
\begin{align}
U&=-\frac{1}{2}\Delta U\sin(2kZ_0) \nonumber \\
\Delta U&= \mathcal{E}_z\mathcal{E}_y\mathcal{E}_x(\mathtt{A}\alpha_{ZY}\mu_{0X}+\mathtt{B}\alpha_{XZ}\mu_{0Y}+\mathtt{C}\alpha_{YX}\mu_{0Z}), \nonumber \nonumber
\end{align}
where $U=U_{J_\tau,m}(\mathbf{R}_0)$ is an effective potential energy surface with depth and sign dictated by $\Delta U=\Delta U_{J_\tau,m}$. The `temperature' $\Delta U/k_B$ serves as a convenient measure of the strength and sign of our chiral optical force.

The reader will note below that the unweighted average of our chiral optical force over rotational states \textit{vanishes}, in accord with the principle of spectroscopic stability \cite{vanVleck32a}. This does not render our chiral optical force trivial, however, as no individual molecule experiences this average force and there are different \textit{numbers} of rotational states with $F_z<0$ and $F_z>0$, giving an (enantioselective) bias in terms of the number of molecules that would be pushed in the $\mp z$ directions for e.g. a thermal distribution of rotational states\footnote{Consideration of rotational states in the $J\in\{0,\dots,10\}$ manifolds should be sufficient for experiments using supersonic expansion beams with rotational temperatures around $1\,\mathrm{K}$.}. To enhance the enantioselectivity, one could selectively populate rotational state(s) with either $F_z<0$ or $F_z>0$ for a given enantiomer (and thus either $F_z>0$ or $F_z<0$ for the opposite enantiomer).

\subsection{Norbornenone}
Let us first consider ($1S$,$4S$)-norbornenone with isotopic constitution ${}^{12}\mathrm{C}_7{}^1\mathrm{H}_8{}^{16}\mathrm{O}$. We take the rotational constants to be $A/2\pi\hbar=3.65\,\mathrm{GHz}$, $B/2\pi\hbar=2.21\,\mathrm{GHz}$, and $C/2\pi\hbar=1.99\,\mathrm{GHz}$ and the chirally sensitive molecular properties to be $\alpha_{ZY}\mu_{0X}=2.76\times10^{-70}\,\mathrm{C}^3\,\mathrm{m}^3\,\mathrm{J}^{-1}$, $\alpha_{XZ}\mu_{0Y}=8.72\times10^{-71}\,\mathrm{C}^3\,\mathrm{m}^3\,\mathrm{J}^{-1}$, and $\alpha_{YX}\mu_{0Z}=-4.85\times 10^{-71}\,\mathrm{C}^3\,\mathrm{m}^3\,\mathrm{J}^{-1}$.

\begin{figure}[h!]
\centering
\includegraphics[width=\textwidth]{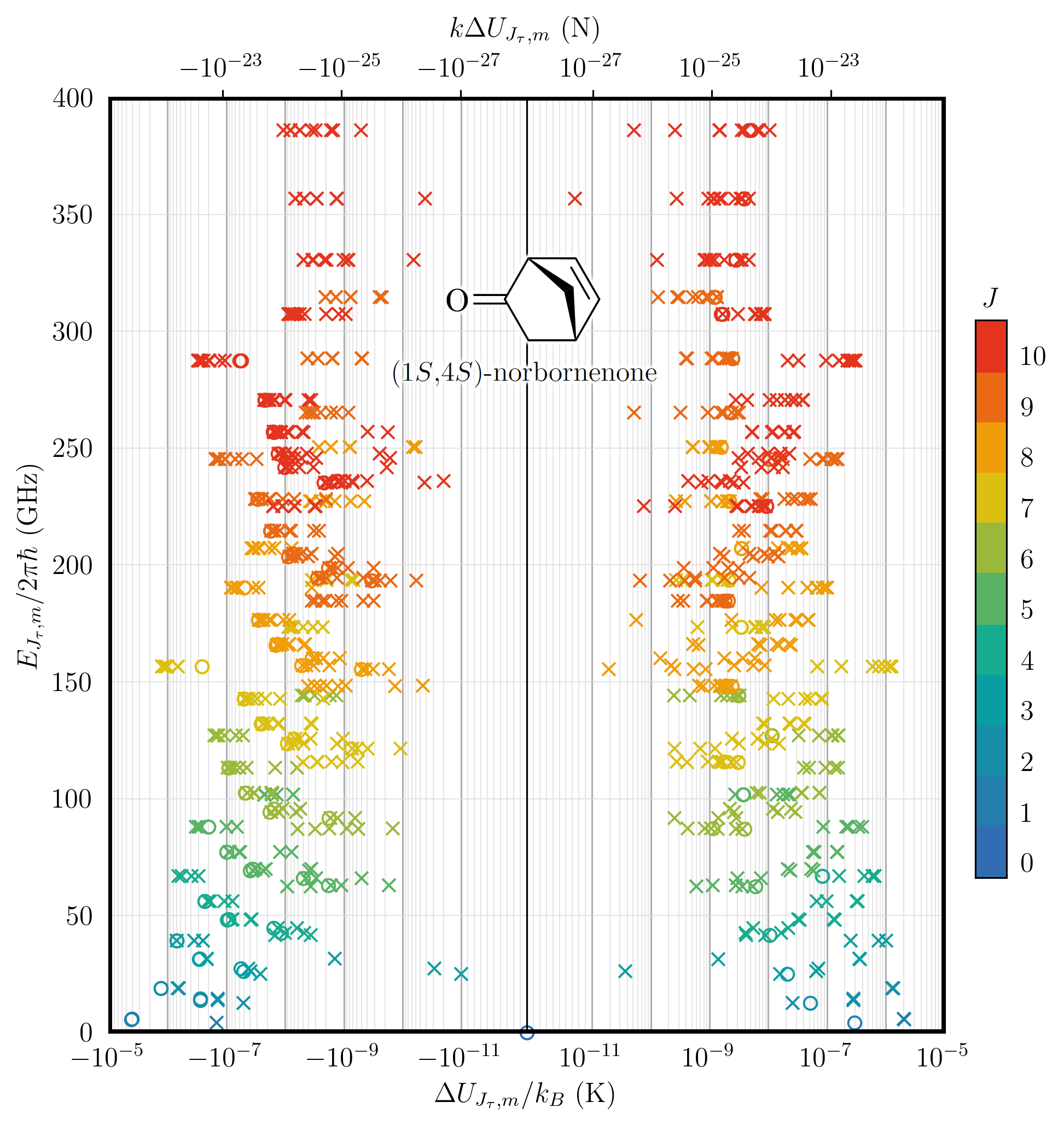}
\caption{\small A scatter plot depicting the strength and sign of our chiral optical force for the rotational states of ($1\textit{S}$,$4\textit{S}$)-norbornenone in the $J\in\{0,\dots,10\}$ manifolds. Each circle corresponds to a rotational state with $m=0$ and each cross corresponds to a pair of rotational states with $m=\pm|m|\ne0$.}
\label{Norbornenone}
\end{figure}

Fig.~\ref{Norbornenone} is a scatter plot depicting the strength and sign of our chiral optical force for rotational states in the $J\in\{0,\dots,10\}$ manifolds, expressed in terms of the `temperature' $\Delta U/k_B$. The strongest force is found for the $|\psi_{1_0,0}\rangle$ rotational state, giving 
\begin{align}
\frac{\Delta U}{k_B}=-4.16\,\mu\mathrm{K} \ \ \ \ \ \ \ \ \ \ (k\Delta U=-3.39\times10^{-22}\,\mathrm{N}). \nonumber
\end{align}
There are $908$ rotational states with $F_z<0$ and only $862$ rotational states with $F_z>0$, giving an overall preference for $F_z<0$. We emphasise here than for a given configuration of the electric field, our chiral optical force has opposite signs for opposite enantiomers in the same rotational state.

\subsection{Isotopically chiral difluorocyclohexane}
Let us now consider the isotopically chiral form of difluorocyclohexane constructed by taking the achiral molecule ($1\textit{R}$,$2\textit{S}$)-$1$,$2$-difluorocyclohexane with isotopic constitution ${}^{12}\mathrm{C}_6{}^1\mathrm{H}_{10}{}^{19}\mathrm{F}_2$ and substituting ${}^{13}\mathrm{C}$ at positions $1$, $5$, and $6$. We take the rotational constants to be $A/2\pi\hbar=2.76\,\mathrm{GHz}$, $B/2\pi\hbar=1.88\,\mathrm{GHz}$, and $C/2\pi\hbar=1.39\,\mathrm{GHz}$ and the chirally sensitive molecular properties to be $\alpha_{ZY}\mu_{0X}=3.75\times10^{-70}\,\mathrm{C}^3\,\mathrm{m}^3\,\mathrm{J}^{-1}$, $\alpha_{XZ}\mu_{0Y}=1.53\times10^{-71}\,\mathrm{C}^3\,\mathrm{m}^3\,\mathrm{J}^{-1}$, and $\alpha_{YX}\mu_{0Z}=2.87\times 10^{-71}\,\mathrm{C}^3\,\mathrm{m}^3\,\mathrm{J}^{-1}$.

\begin{figure}[h!]
\centering
\includegraphics[width=\textwidth]{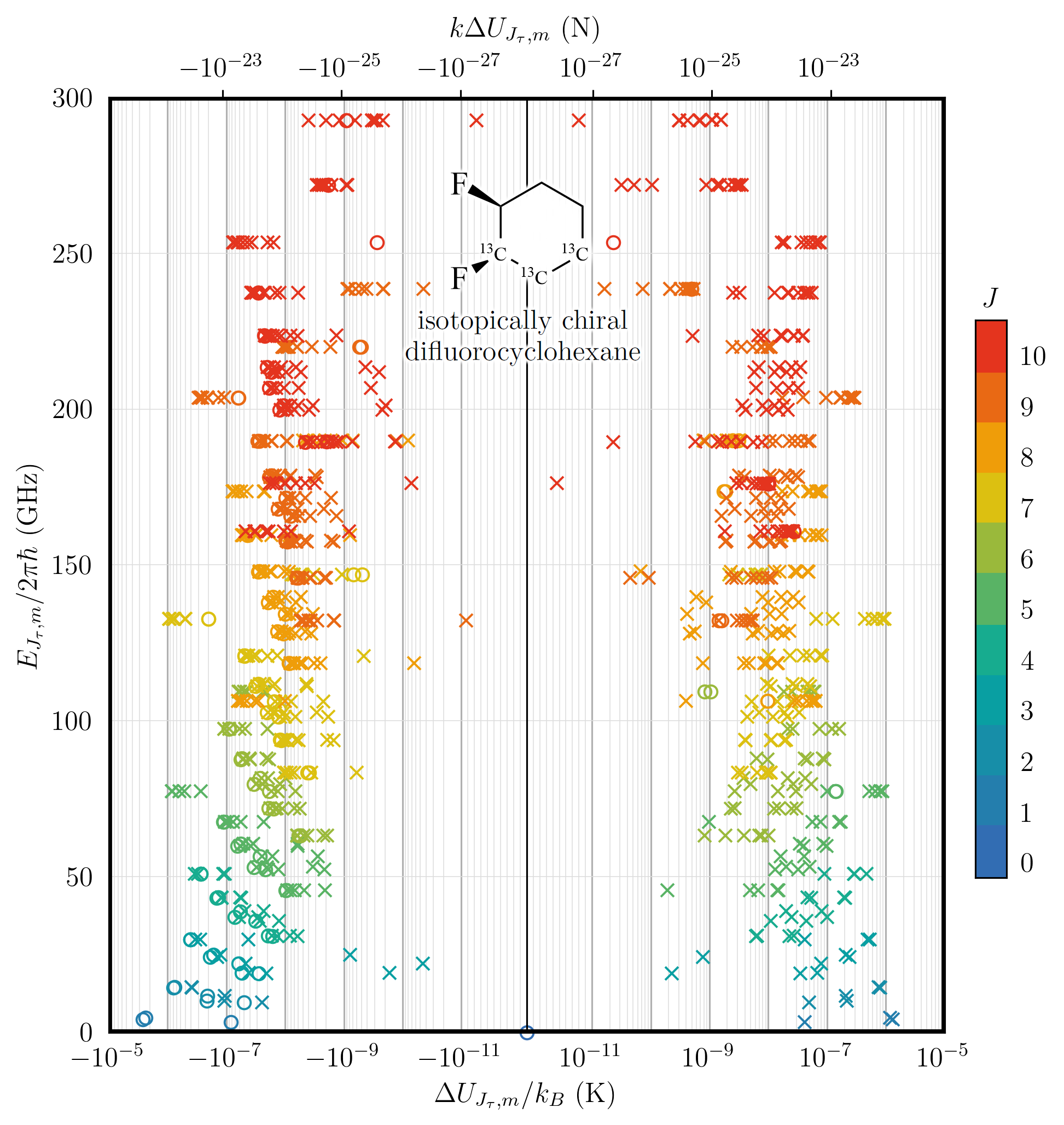}
\caption{\small A scatter plot depicting the strength and sign of our chiral optical force for the rotational states of an isotopically chiral form of difluorocyclohexane in the $J\in\{0,\dots,10\}$ manifolds. Each circle corresponds to a rotational state with $m=0$ and each cross corresponds to a pair of rotational states with $m=\pm|m|\ne0$.}
\label{Difluorocyclohexane}
\end{figure}

Fig.~\ref{Difluorocyclohexane} is a scatter plot depicting the strength and sign of our chiral optical force for rotational states in the $J\in\{0,\dots,10\}$ manifolds. The strongest force is again found for the $|\psi_{1_0,0}\rangle$ rotational state, giving 
\begin{align}
\frac{\Delta U}{k_B}&=-2.64\,\mu\mathrm{K} \ \ \ \ \ \ \ \ \ \ (k\Delta U=-2.15\times10^{-22}\,\mathrm{N}). \nonumber
\end{align}
There are $951$ rotational states with $F_z<0$ and only $819$ rotational states with $F_z>0$, giving an overall preference for $F_z<0$.

\subsection{Comparison with other chiral optical forces}
It is instructive to compare the results above for our chiral optical force with the corresponding results for other chiral optical forces proposed to date. 

In \cite{Cameron14b}, for example, we showed that an optical helicity lattice can exert a chiral optical force on a small chiral molecule of the form
\begin{align}
\mathbf{F}&\approx\frac{1}{c}2k\sin\vartheta  \mathcal{E}_\|\mathcal{E}_\bot\cos^2\vartheta G^\prime\cos(2k\sin\vartheta Z_0)\hat{\mathbf{z}},
 \label{HelicityF}
\end{align}
where $2\vartheta$ is the angular separation of the waves comprising the lattice, $\mathcal{E}_\|$ dictates the amplitude and phase of the parallel polarized wave, $\mathcal{E}_\bot$ dictates the amplitude and phase of the perpendicularly polarized wave, and $G^\prime$ is the imaginary part of the molecule's vibronic optical activity polarizability pseudoscalar \cite{Barron04a}. To compare the strength of our chiral optical force with the strength of this helicity-based force, we consider a dimensionless ratio $\mathcal{R}=\mathcal{R}_{J_\tau,m}$ given by
\begin{align}
\mathcal{R}&=\left|\frac{k\mathcal{E}_z\mathcal{E}_y\mathcal{E}_x(\mathtt{A}\alpha_{ZY}\mu_{0X}+\mathtt{B}\alpha_{XZ}\mu_{0Y}+\mathtt{C}\alpha_{YX}\mu_{0Z})}{2k\sin\vartheta  \mathcal{E}_\|\mathcal{E}_\bot\cos^2\vartheta G^\prime/c}\right|, \nonumber
\end{align}
where the numerator follows from Eq.~(\ref{F}) and the denominator follows from Eq.~(\ref{HelicityF}). 

For the helicity lattice, we consider an angular separation of $2\vartheta=60.0^\circ$ and field strengths of $\mathcal{E}_\|=\mathcal{E}_\bot=3.16\,\mathrm{kV}\,\mathrm{mm}^{-1}$, matching the field strengths $\mathcal{E}_y=\mathcal{E}_x=3.16\,\mathrm{kV}\,\mathrm{mm}^{-1}$ of the standing wave to give a fair comparison.

For ($1S$,$4S$)-norbornenone, we estimate the optical activity polarizability to be $G^\prime=3\times 10^{-36}\,\mathrm{m}\,\mathrm{kg}^{-1}\,\mathrm{s}^3\,\mathrm{A}^2$ from vapour-phase measurements \cite{Lahiri14a} together with an appropriate frequency scaling \cite{Cameron14c}. Norbornenone is exceptionally well suited to the helicity-based force, as $G^\prime$ is around two orders of magnitude larger than for comparably sized organic molecules \cite{Lahiri14a}. Even so, we find that our chiral optical force can be several orders of magnitude larger than the helicity-based force. For the $|\psi_{1_0,0}\rangle$ rotational state, we obtain a ratio of
\begin{align}
\mathcal{R}&\sim 10^3. \nonumber
\end{align}
For other molecules with more typical values of $G^\prime$, ratios of 
\begin{align}
\mathcal{R}&\sim 10^5 \nonumber
\end{align}
or higher are possible. That is to say, a disparity in strength of \textit{five} orders of magnitude or more. Even higher ratios can be found at lower frequencies, as $\mathcal{R}\propto 1/\omega$ far off resonance. 

Isotopically chiral molecules are poorly suited to the helicity-based force, as the optical activity pseudoscalar $G^\prime$ is usually very small, its electronic contributions formally vanishing in the Born-Oppenheimer approximation \cite{Bunker05a}. Isotopically chiral molecules are nevertheless amenable to our chiral optical force, as shown above for an isotopically chiral form of difluorocyclohexane.

Such comparisons demonstrate that potential applications identified elsewhere for other chiral optical forces \cite{Kucirka96a, Canaguier-Durand13a, Cameron14b, Cameron14c, Bradshaw15a, Rukhlenko16a, Marichez19a, Fang21a, Kakkanattu21a, Stickler21a, Isaule22a, Genet22a, Okamoto22a, Forbes22a} should be enormously more robust and easier to implement using our chiral optical force instead, as claimed in Sec. \ref{Introduction}. 

\subsection{Relation to other chiroptical phenomena}
Most chiroptical phenomena rely on weak magnetic-dipole and electric-quadrupole interactions \cite{Lough02a, Barron04a}. We emphasise here that our chiral optical force can be orders of magnitude stronger than other chiral optical forces proposed to date because it is based instead on electric-dipole interactions at leading order.

Other chiroptical phenomena based on electric-dipole interactions at leading order include second-order nonlinear sum- and difference-frequency generation \cite{Giordmaine65a, Rentzepis66a} and extensions thereof using a static electric field \cite{Buckingham98a, Fischer03a}, photoelectron circular dichroism \cite{Ritchie76a, Bowering01a}, second-harmonic generation circular dichroism from chiral surfaces \cite{Petralli-Mallow93a, Byers94a}, Rayleigh and Raman optical activity from chiral surfaces \cite{Hecht94a}, Coulomb-explosion imaging \cite{Kitamura01a, Pitzer13a, Herwig13a}, chiral microwave three-wave mixing \cite{Hirota12a, Nafie13a, Patterson13a}, photoexcitation circular dichroism and photoexcitation-induced electron circular dichroism \cite{Beaulieu18a}, chiral high harmonic generation in the electric-dipole approximation \cite{Ayuso19a, Ayuso22a}, ultrafast optical rotation \cite{Ayuso21a}, and anomalous circularly polarized light emission \cite{Wan23a}. Our chiral optical force has elements in common with many of these, in particular its crucial dependence on orientational effects and its nonlinear character. For an incisive perspective on the current ``electric-dipole revolution in chiral measurements'', see \cite{Ayuso22b}.

%
%

\section{Summary and Outlook}
\label{Summary and Outlook}
Taking our lead from nature, we have identified a chiral optical force for small chiral molecules that can be several orders of magnitude stronger than other chiral optical forces proposed to date. Our chiral optical force should render potential applications enormously more robust and easier to implement than previously thought possible. 

There is much still to be done. For a more accurate description of our chiral optical force, the molecule's nuclear spins should be considered explicitly. A perturbative treatment of our chiral optical force like the one presented in this paper should be sufficient for chiral molecular matter-wave interferometry \cite{Cameron14b, Cameron14c, Stickler21a} and the distillation of chiral molecules in chiral optical lattices \cite{Isaule22a}, for example. It remains for us to develop a non-perturbative treatment for more demanding potential applications, including the resolution of enantiomers \cite{Kucirka96a, Canaguier-Durand13a, Cameron14b, Bradshaw15a, Rukhlenko16a, Fang21a, Forbes22a}.

We note here that many different electromagnetic fields can give rise to chiral optical forces based on electric-dipole interactions at leading order. A key ingredient is the use of three mutually orthogonal electric field components, which can be arranged in either a left-handed or a right-handed configuration. These chiral optical forces might be referred to as different ``blends'' of COFFEEE, the chiral optical force identified in this paper being the ``static/lin$\perp$lin blend''. Synthetic chiral light \cite{Ayuso19a} and chiral topological light \cite{Mayer23a} might offer additional possibilities for chiral optical forces based on electric-dipole interactions.

We will return to these and related ideas elsewhere.

%
%

\begin{backmatter}
\bmsection{Funding}
The Royal Society (URF$\backslash$R1$\backslash$191243). 

\bmsection{Acknowledgments}
Robert P. Cameron is a Royal Society University Research Fellow and gratefully acknowledges the Royal Society's support.

\bmsection{Disclosures}
The authors declare no conflicts of interest.

\bmsection{Data availability}
No data were generated or analyzed in the presented research.


\end{backmatter}

%
%

%
%

\begin{appendix}

%
%

\section{Asymmetric rigid rotor}
\label{Asymmetric rigid rotor}
The quantum-mechanical description of the asymmetric rigid rotor has been covered extensively elsewhere \cite{Wang29a, Townes75a, Bernath05a, Bunker05a}. We use the conventions below, which are slightly unusual as we consider quantization along the $y$ axis rather than the $z$ axis.

We take the molecule-fixed coordinates $X$, $Y$, and $Z$ to be arranged in a $\mathrm{III}^r$ configuration with direction cosines $\ell_{aA}$ given by
\begin{align}
\ell_{xX}&=\cos\theta\cos\phi\cos\chi-\sin\phi\sin\chi, \nonumber \\
\ell_{xY}&=-\cos\theta\cos\phi\sin\chi-\sin\phi\cos\chi, \nonumber \\
\ell_{xZ}&=\sin\theta\cos\phi, \nonumber \\
\ell_{yX}&=-\sin\theta\cos\chi, \nonumber \\
\ell_{yY}&=\sin\theta\sin\chi, \nonumber \\
\ell_{yZ}&=\cos\theta, \nonumber \\
\ell_{zX}&=-\cos\theta\sin\phi\cos\chi-\cos\phi\sin\chi, \nonumber \\
\ell_{zY}&=\cos\theta\sin\phi\sin\chi-\cos\phi\cos\chi, \nonumber \\
\ell_{zZ}&=-\sin\theta\sin\phi, \nonumber
\end{align}
where $0\le\theta\le\pi$, $0\le\phi<2\pi$, and $0\le\chi<2\pi$ are Euler angles relating $X$, $Y$, and $Z$ to the laboratory-fixed coordinates $x$, $y$, and $z$. The unperturbed rotational Hamiltonian is thus
\begin{align}
H^{(0)}&=\frac{1}{\hbar^2}(A J^2_X+B J^2_Y+C J^2_Z) \nonumber 
\end{align}
with
\begin{align}
J_X&=-\mathrm{i}\hbar\left(\sin\chi\frac{\partial}{\partial\theta}-\csc\theta\cos\chi\frac{\partial}{\partial\phi}+\cot\theta\cos\chi\frac{\partial}{\partial\chi}\right), \nonumber \\
J_Y&=-\mathrm{i}\hbar\left(\cos\chi\frac{\partial}{\partial\theta}+\csc\theta\sin\chi\frac{\partial}{\partial\phi}-\cot\theta\sin\chi\frac{\partial}{\partial\chi}\right), \nonumber \\
J_Z&=-\mathrm{i}\hbar\frac{\partial}{\partial\chi}, \nonumber
\end{align}
where $A>B>C$ are the molecule's equilibrium rotational constants and $\mathbf{J}$ is the molecule's rotational angular momentum.

We take the unperturbed rotational energy eigenstates $|J_{\tau},m\rangle^{(0)}$ and associated energy eigenvalues $E_{J_\tau,m}^{(0)}$ to satisfy
\begin{align}
H^{(0)}|J_\tau,m\rangle^{(0)}&=E_{J_\tau,m}^{(0)}|J_\tau,m\rangle^{(0)}, \nonumber \\
(J_X^2+J_Y^2+J_Z^2)|J_\tau,m\rangle^{(0)}&=\hbar^2J(J+1)|J_\tau,m\rangle^{(0)} ,\nonumber \\
J_y|J_\tau,m\rangle^{(0)}&=\hbar m|J_\tau,m\rangle^{(0)} \nonumber
\end{align}
with
\begin{align}
J_x&=-\mathrm{i}\hbar\left(-\sin\phi\frac{\partial}{\partial\theta}-\cot\theta\cos\phi\frac{\partial}{\partial\phi}+\csc\theta\cos\phi\frac{\partial}{\partial\chi}\right), \nonumber \\
J_y&=-\mathrm{i}\hbar\frac{\partial}{\partial\phi}, \nonumber \\
J_z&=-\mathrm{i}\hbar\left(-\cos\phi\frac{\partial}{\partial\theta}+\cot\theta\sin\phi\frac{\partial}{\partial\phi}-\csc\theta\sin\phi\frac{\partial}{\partial\chi}\right), \nonumber
\end{align}
where $J\in\{0,1,\dots\}$ is the rotational angular momentum quantum number, $\tau\in\{-J,\dots,J\}$ is a label that increases with increasing energy and $m\in\{-J,\dots,J\}$ is the laboratory-fixed rotational angular momentum projection quantum number.

To help us identify the unperturbed rotational energy eigenstates $|J_{\tau},m\rangle^{(0)}$ and associated energy eigenvalues $E_{J_\tau,m}^{(0)}$ explicitly, we work in a basis of unperturbed symmetric rigid rotor states $|J,\kappa,m\rangle^{(0)}$ satisfying
\begin{align}
(J_X^2+J_Y^2+J_Z^2)|J,\kappa,m\rangle^{(0)}&=\hbar^2 J(J+1)|J,\kappa,m\rangle^{(0)}, \nonumber \\
J_Z|J,\kappa,m\rangle^{(0)}&=\hbar\kappa|J,\kappa,m\rangle^{(0)}, \nonumber \\
J_y|J,\kappa,m\rangle^{(0)}&=\hbar m|J,\kappa,m\rangle^{(0)}, \nonumber
\end{align}
where $\kappa\in\{-J,\dots,J\}$ is the molecule-fixed rotational angular momentum projection quantum number, the corresponding wavefunctions $\langle\theta,\phi,\chi|J,\kappa,m\rangle^{(0)}$ being given by
\begin{align}
\langle \theta,\phi,\chi|J,\kappa,m\rangle^{(0)}&=\sqrt{\frac{(J+m)!(J-m)!(J+\kappa)!(J-\kappa)!(2J+1)}{8\pi^2}} \nonumber \\
&
\times\sum_{\sigma=\mathrm{max}(0,\kappa-m)}^{\mathrm{min}(J-m,J+\kappa)}(-1)^\sigma\frac{(\cos\frac{1}{2}\theta)^{2J+\kappa-m-2\sigma}(-\sin\frac{1}{2}\theta)^{m-\kappa+2\sigma}}{\sigma!(J-m-\sigma)!(m-\kappa+\sigma)!(J+\kappa-\sigma)!}\mathrm{e}^{\mathrm{i}m\phi}\mathrm{e}^{\mathrm{i}\kappa\chi}. \nonumber
\end{align}
The $|J,\kappa,m\rangle^{(0)}$ render the unperturbed rotational Hamiltonian $H^{(0)}$ block diagonal in $J$, as
\begin{align}
{}^{(0)}\langle J,\kappa,m|H^{(0)}|J^\prime,\kappa^\prime,m^\prime\rangle^{(0)}&=\delta_{JJ^\prime}\delta_{mm^\prime}\Bigg(\frac{1}{2}(A+B)J(J+1)\delta_{\kappa\kappa^\prime}+\left[C-\frac{1}{2}(A+B)\right]k^2\delta_{\kappa\kappa^\prime} \nonumber \\
&+\frac{1}{4}(A-B) \nonumber \\
&\times\Big\{\sqrt{[J(J+1)-(\kappa+1)\kappa][J(J+1)-(\kappa+2)(\kappa+1)]}\delta_{\kappa\kappa^\prime-2} \nonumber \\
&+\sqrt{[J(J+1)-(\kappa-1)\kappa][J(J+1)-(\kappa-2)(\kappa-1)]}\delta_{\kappa\kappa^\prime+2}\Big\}\Bigg). \nonumber
\end{align}
The diagonalization of $H^{(0)}$ can completed analytically for $J\in\{0,\dots,5\}$, giving
\begin{align}
|0_0,0\rangle^{(0)}&=|0,0,0\rangle^{(0)}, \nonumber \\
|1_{-1},m\rangle^{(0)}&=\frac{1}{\sqrt{2}}(|1,1,m\rangle^{(0)}-|1,-1,m\rangle^{(0)}), \nonumber \\
|1_0,m\rangle^{(0)}&=\frac{1}{\sqrt{2}}(|1,1,m\rangle^{(0)}+|1,-1,m\rangle^{(0)}), \nonumber \\
|1_1,m\rangle^{(0)}&=|1,0,m\rangle^{(0)} \nonumber
\end{align}
with
\begin{align}
E_{0_0,m}^{(0)}&=0, \nonumber \\
E_{1_{-1},m}^{(0)}&=B+C, \nonumber \\
E_{1_0,m}^{(0)}&=A+C, \nonumber \\
E_{1_1,m}^{(0)}&=A+B, \nonumber 
\end{align}
for example. The diagonalization of $H^{(0)}$ must be completed numerically for $J\in\{6,\dots\}$.

%
%

\section{Coefficients}
\label{Coefficients}
The coefficients $\mathtt{a}=\mathtt{a}_{J_\tau,m}$, $\mathtt{b}=\mathtt{b}_{J_\tau,m}$, $\mathtt{c}=\mathtt{c}_{J_\tau,m}$, $\mathtt{A}=\mathtt{A}_{J_\tau,m}$, $\mathtt{B}=\mathtt{B}_{J_\tau,m}$, and $\mathtt{C}=\mathtt{C}_{J_\tau,m}$ are given by
\begin{align}
\mathtt{a}&={}^{(0)}\langle J_\tau,m|\left(\ell_{yZ}\ell_{xY}\ell_{zX}+\ell_{yY}\ell_{xZ}\ell_{zX}\right)|J_\tau,m\rangle^{(0)}, \label{aCoefficient} \\
\mathtt{b}&={}^{(0)}\langle J_\tau,m|\left(\ell_{yX}\ell_{xZ}\ell_{zY}+\ell_{yZ}\ell_{xX}\ell_{zY}\right)|J_\tau,m\rangle^{(0)}, \label{bCoefficient} \\
\mathtt{c}&={}^{(0)}\langle J_\tau,m|\left(\ell_{yY}\ell_{xX}\ell_{zZ}+\ell_{yX}\ell_{xY}\ell_{zZ}\right)|J_\tau,m\rangle^{(0)}, \label{cCoefficient} \\
\mathtt{A}&=\sum_{J^\prime=0}^\infty\sum_{\tau^\prime=-J^\prime}^{J^\prime}\sum_{m^\prime=-J^\prime}^{J^\prime}\frac{2}{E_{J^\prime_{\tau^\prime},m^\prime}^{(0)}-E_{J_\tau,m}^{(0)}} \nonumber \\
&\times\Re[{}^{(0)}\langle J_\tau,m|(\ell_{yZ}\ell_{xY}+\ell_{yY}\ell_{xZ})|J^\prime_{\tau^\prime},m^\prime\rangle^{(0)}{}^{(0)}\langle J^\prime_{\tau^\prime},m^\prime|\ell_{zX}|J_\tau,m\rangle^{(0)}], \label{ACoefficient} \\
\mathtt{B}&=\sum_{J^\prime=0}^\infty\sum_{\tau^\prime=-J^\prime}^{J^\prime}\sum_{m^\prime=-J^\prime}^{J^\prime}\frac{2}{E_{J^\prime_{\tau^\prime},m^\prime}^{(0)}-E_{J_\tau,m}^{(0)}} \nonumber \\
&\times\Re[{}^{(0)}\langle J_\tau,m|(\ell_{yX}\ell_{xZ}+\ell_{yZ}\ell_{xX})|J^\prime_{\tau^\prime},m^\prime\rangle^{(0)}{}^{(0)}\langle J^\prime_{\tau^\prime},m^\prime|\ell_{zY}|J_\tau,m\rangle^{(0)}] ,\label{BCoefficient} \\
\mathtt{C}&=\sum_{J^\prime=0}^\infty\sum_{\tau^\prime=-J^\prime}^{J^\prime}\sum_{m^\prime=-J^\prime}^{J^\prime}\frac{2}{E_{J^\prime_{\tau^\prime},m^\prime}^{(0)}-E_{J_\tau,m}^{(0)}} \nonumber \\
&\times\Re[{}^{(0)}\langle J_\tau,m| 
(\ell_{yY}\ell_{xX}+\ell_{yX}\ell_{xY})|J^\prime_{\tau^\prime},m^\prime\rangle^{(0)}{}^{(0)}\langle J^\prime_{\tau^\prime},m^\prime|\ell_{zZ}|J_\tau,m\rangle^{(0)}], \label{CCoefficient}
\end{align}
where terms with $|J^\prime_{\tau^\prime},m^\prime\rangle^{(0)}=|J_\tau,m\rangle^{(0)}$ are to be excluded from the summations. Eqs.~(\ref{aCoefficient})-(\ref{CCoefficient}) can be evaluated in closed form for rotational states in the $J\in\{0,1,2,3,4\}$ manifolds, as the unperturbed rotational energy eigenstates $|J^\prime_{\tau^\prime},m^\prime\rangle^{(0)}$ and associated energy eigenvalues $E_{J^\prime_{\tau^\prime},m^\prime}^{(0)}$ of importance can be found analytically. For $J=0$, we obtain $\mathtt{a}=\mathtt{b}=\mathtt{c}=\mathtt{A}=\mathtt{B}=\mathtt{C}=0$, which is unsurprising given that $|0_0,0\rangle^{(0)}$ has isotropic character whereas our chiral optical force depends crucially on orientational effects. For $J=1$, we obtain the results presented in Table \ref{LowCoefficients}, where it can be seen that $\mathtt{a}$, $\mathtt{b}$, $\mathtt{c}$, $\mathtt{A}$, $\mathtt{B}$, and $\mathtt{C}$ depend on the magnitude but not the sign of $m$ and that they satisfy relationships like
\begin{align}
\sum_{m=-J}^J\mathtt{A}_{J_\tau,m}&=0, \nonumber
\end{align}
in accord with the principle of spectroscopic stability \cite{vanVleck32a}. Eqs.~(\ref{aCoefficient})-(\ref{CCoefficient}) cannot be evaluated in closed form for rotational states in the $J\in\{5,\dots\}$ manifolds, as some or all of the $|J^\prime_{\tau^\prime},m^\prime\rangle^{(0)}$ and  $E_{J^\prime_{\tau^\prime},m^\prime}^{(0)}$ of importance need to be found numerically.

\begin{table}[h!]
\begin{tabular}{c|c|c|c|c}
& $|\psi_{1_{-1}, m}\rangle$ & $|\psi_{1_0, m}\rangle$ & $|\psi_{1_1, m}\rangle$ \\ \hline
$\mathtt{a}$ & $0$  & $\frac{(3|m|-2)}{10}$ & $\frac{(2-3|m|)}{10}$ \\
$\mathtt{b}$ & $\frac{(2-3|m|)}{10}$ & $0$ & $\frac{(3|m|-2)}{10}$ \\
$\mathtt{c}$ & $\frac{(3|m|-2)}{10}$ & $\frac{(2-3|m|)}{10}$ &  $0$ \\
$\mathtt{A}$ & $\frac{(2-3|m|)(B-C)}{5[3(B-C)^2-8A(B+C)]}$ & $\frac{(3|m|-2)}{20}\left[\frac{3}{(B-C)}+\frac{1}{(B+3C)}\right]$ & $\frac{(2-3|m|)}{20}\left[\frac{3}{(C-B)}+\frac{1}{(C+3B)}\right]$  \\
$\mathtt{B}$ & $\frac{(2-3|m|)}{20}\left[\frac{3}{(A-C)}+\frac{1}{(A+3C)}\right]$ & $\frac{(2-3|m|)(C-A)}{5[3(C-A)^2-8B(A+C)]}$ & $\frac{(3|m|-2)}{20}\left[\frac{3}{(C-A)}+\frac{1}{(C+3A)}\right]$ \\
$\mathtt{C}$ & $\frac{(3|m|-2)}{20}\left[\frac{3}{(A-B)}+\frac{1}{(A+3B)}\right]$ & $\frac{(2-3|m|)}{20}\left[\frac{3}{(B-A)}+\frac{1}{(B+3A)}\right]$ & $\frac{(2-3|m|)(A-B)}{5[3(A-B)^2-8C(A+B)]}$ \\
\end{tabular}
\caption{\small The coefficients $\mathtt{a}$, $\mathtt{b}$, $\mathtt{c}$, $\mathtt{A}$, $\mathtt{B}$, and $\mathtt{C}$ for rotational states in the $J=1$ manifold.}
\label{LowCoefficients}
\end{table}

\end{appendix}
 
%
%

\end{document}